\documentclass[cmp]{swjour}  
\usepackage{amsmath}
\usepackage{amsfonts,amssymb}
\usepackage{times}
\spnewtheorem{remarks}{Remarks}{\bf}{\rm}

\spnewtheorem{remark}{Remark}{\bf}{\rm}
\numberwithin{theorem}{section}
\numberwithin{proposition}{section}
\numberwithin{corollary}{section}
\numberwithin{lemma}{section}
\numberwithin{definition}{section}
\numberwithin{remark}{section}
\numberwithin{remarks}{section}
\numberwithin{equation}{section}

\renewcommand{\P}{\mathbb{P}}
\newcommand{\C}{\mathbb{C}}
\newcommand{\T}{\mathbb{T}}
\newcommand{\G}{\mathbb{G}}
\newcommand{\R}{\mathbb{R}}
\let\Im\undefined
\let\Re\undefined
\DeclareMathOperator{\Im}{Im}
\DeclareMathOperator{\Re}{Re}

\makeatletter
\newcounter{numcount}
\newcommand{\labelnummer}{\mbox{(\roman{numcount})}}%
\newenvironment{indentnummer}%
        {\let\curlabelspeicher\@currentlabel%
         \begin{list}{\labelnummer}{\usecounter{numcount}%
    \topsep1ex\partopsep2ex\parsep0pt\itemsep1ex\@plus1\p@%
                      \labelwidth2.5em\itemindent0em\labelsep1em%
                      \leftmargin3em}%
         \let\saveitem\item%
         \def\item{\saveitem%
                   \def\@currentlabel{\curlabelspeicher\labelnummer}%
                   \let\label\bemlabel}}%
       {\end{list}}%
\newenvironment{indentnummer*}%
        {\begin{list}{\labelnummer}{\usecounter{numcount}%
   \topsep1ex\partopsep2ex\parsep0pt\itemsep1ex
  \labelwidth2.5em\itemindent0em\labelsep1em%
       \leftmargin3.5em
                      }}%
       {\end{list}}%
\newenvironment{nummer}%
        {\let\curlabelspeicher\@currentlabel%
   \begin{list}{\labelnummer}{\usecounter{numcount}\leftmargin0em%
\topsep1ex\partopsep2ex\parsep0pt\itemsep0.5ex
       \labelwidth2.5em\itemindent3em\labelsep1em}%
         \let\saveitem\item%
         \def\item{\saveitem%
        \def\@currentlabel{\curlabelspeicher\labelnummer}%
                   \let\label\bemlabel}}%
       {\end{list}}%
\def\itemref#1{\expandafter\@setref\csname r@#1item\endcsname\@firstoftwo{#1}}%
\def\bemlabel#1{\@bsphack%
      \protected@write\@auxout{}%
             {\string\newlabel{#1}{{\@currentlabel}{\thepage}}}%
      \ifmmode\else%
      \protected@write\@auxout{}%
     {\string\newlabel{#1item}{{\labelnummer}{\thepage}}}%
      \fi%
      \@esphack}%
\makeatother
       
\begin{document}
\title{Absolutely Continuous Spectra of Quantum Tree Graphs
with Weak Disorder}
\titlerunning{Quantum Tree Graphs with Disorder}
\author{Michael Aizenman\inst{1} \and Robert Sims\inst{2}
\and Simone Warzel\inst{1}\footnote{On leave from: Institut f\"ur Theoretische Physik, Universit\"at Erlangen-N\"urnberg, Germany.}}
\institute{Departments of Mathematics and Physics,
            Princeton University, Princeton NJ 08544, USA \and
Department of Mathematics, University of California at Davis,
Davis CA 95616, USA  }
\authorrunning{Aizenman, Sims, Warzel}
\date{Version: June 3, 2005}
\communicated{}
\maketitle  
\begin{abstract}
We consider the Laplacian on a rooted metric tree graph with
branching number
$ K \geq 2 $ and random edge lengths given by independent and
identically distributed bounded variables.
Our main result is the stability of the absolutely continuous
spectrum for weak disorder.  
A useful tool in the discussion is a function  which expresses 
a directional transmission amplitude to infinity and forms a 
generalization of the Weyl-Titchmarsh function to trees.  
The proof of the main result rests on upper bounds on the  range of 
fluctuations of this quantity in the limit of weak disorder.   
\end{abstract}
\tableofcontents
\section{Introduction}
A quantum graph (QG) is a metric graph with an associated Laplace-like operator
acting on the $L^2$-space of the union of the graph edges.
The spectral and dynamical properties of such operators have been of
interest both
because this model mimics situations realizable with quantum dots and
wires, and because QGs may provide a simple setup
elucidating issues which are also of relevance for Schr\"odinger
operators and Laplacians on manifolds (see \cite{KoSm99,SchSmil00,Kuch02,KosSch04}
and references therein).  Examples of such topics are the
Gutzwiller trace formula and the transition associated with the
spectral and dynamical localization due to disorder.
The main results of this work pertain to quantum {\em tree graphs}
whose edge lengths are randomly stretched, but remain close to a common value.
The goal is to present new results concerning the persistence
of absolutely continuous spectra under weak disorder.   A secondary
goal is to demonstrate, in the QG context, a new technique for the
proof of absolutely continuous spectrum which is also effective
for discrete random Schr\"odinger operators on trees as was proven in \cite{ASW05}.
\subsection{Random quantum trees and their spectra}
A rooted metric tree graph $\T$ with branching number $K$
consists, for us, of a countably infinite set of vertices,
one of which is being labeled as the root, $0 $,
and a set
$ \mathcal{E} $ of edges, each joining a pair of
vertices, such that:
{\it 1.\/}  the graph is edge connected,
{\it 2.\/} there are no closed loops,
{\it 3.\/}  each vertex has $K+1$ edges except for the root which has
only one edge.
Each edge $ e \in \mathcal{E}$ is assigned a positive finite length
$ L_e \in (0, \infty) $ and is parametrized by a variable with values in
$[0, L_e]$. Thus, the union of
the edges has the natural coordinates $ l \in [0,L_e]$. The orientation
for the latter is chosen so that  $ l $ increases away from
the root, and we denote by $ {}' $ the derivative with respect to those coordinates.

Our discussion concerns the spectral properties of the Laplacian
\begin{equation}\label{eq:Delta}
	- \Delta_\T \, \psi_e =  - \psi_e''\, ,
\end{equation}
which acts in the Hilbert space
${\rm L}^2(\T)= \bigoplus_{e \in \mathcal{E}}\, {\rm
L}^2 [0,L_e] $ of complex-valued square-integrable functions
$ \psi = \oplus_{e \in \mathcal{E}} \, \psi_e $ defined over the
union of the graph edges.
The Laplacian is rendered essentially self-adjoint through the
imposition of boundary
conditions (BC) on the functions in its domain; here we take these
to be the Kirchhoff
conditions at internal vertices and $ \alpha $-BC at the root. More precisely, the domain consists
of functions such that $ \psi_e \in {\rm H^2}[0,L_e] $ for all $ e
\in \mathcal{E} $ and
\begin{enumerate}
\item at each vertex $\psi $ is continuous.
\item at internal vertices the net flux defined by the directional
derivatives vanishes, i.e.,
\begin{equation}
            \psi_e'(L_e) = \sum_{f \in \mathcal{N}^+_e }\psi_{f}'(0) \label{eq:Kirch}
\end{equation}
where $ \mathcal{N}^+_e $ is the collection of edges which
are forward to $ e $ as seen from the root.
\item at the root
\begin{equation}\label{eq:root}
\cos(\alpha) \, \psi_0(0) - \sin(\alpha) \, \psi_0'(0)  = 0
\end{equation}
with some $ \alpha \in [0,\pi) $.
\end{enumerate}
An extensive discussion of other boundary conditions which yield  
self adjointness can be found in \cite{KosSch99,Kuch04}.  Among those  
is the class of symmetric BC; the adaptation of the argument to this case 
is discussed in Section~\ref{sec:extensions}.

\subsection{Statement of the main result}
Our discussion will focus on the absolutely continuous (AC) component of the
spectrum of the Laplacian on deformed metric trees.
Before presenting the main result let us note the following fact, which
may, for instance, be deduced from Theorem~\ref{thm:spec} in Appendix~\ref{app:more}.
\begin{proposition}\label{prop:Sol}
The AC spectrum of $ - \Delta_\T $ is independent of the boundary condition  at the root, i.e., of $ \alpha
\in [0,\pi) $.
\end{proposition}
For the regular tree $ \T{} $ with constant edge lengths $ L  \in
(0, \infty) $
and branching number $ K \in \mathbb{N} $ one has \cite[Example~6.3]{Sol04}
\begin{equation}\label{eq:specSol}
\sigma_{\rm ac}(-\Delta_{\T{}})  =
\bigcup_{n = 0}^\infty \left[ \left(\frac{\pi n + \theta
}{L}\right)^2 ,\left(\frac{\pi (n+1) - \theta}{L}\right)^2 \right]
\end{equation}
where $ \theta := \arctan\left[\left( K^{1/2} - K^{-1/2}\right)/2\right] $.
In particular, this implies that the AC spectrum of $ - \Delta_{\T{}} $
has band structure if $ K \geq 2 $. As an aside, we note that for $ K \geq 2 $
there occur infinitely degenerate eigenvalues in the band gaps \cite{Sol04}. \\
The main object of interest in this paper is the AC spectrum of the
Laplacian on
random deformations of $ \T $.
\begin{definition}\label{def:def}
A \emph{random deformation}  $\T({\lambda,\omega}) $
of the regular rooted metric tree $ \T{} $ is a rooted metric tree
graph, which has the same
vertex set and neighboring relations as $ \T{} $, but the edge lengths
are given by
\begin{equation}
L_e(\lambda,\omega) := L \, \exp\left( \lambda \, \omega_e\right)
\end{equation}
with a collection of real-valued, independent, and identically
distributed (iid) bounded random variables $ \omega = \left\{ \omega_e \right\}_{e \in \mathcal{E}} $. 
The parameter $ \lambda \in[0,1]  $
controls the strength of the disorder and $ L > 0 $ stands for the
edge length of $ \T{} $.
\end{definition}
Our main result is
\begin{theorem}\label{thm:main}
For a random deformation, $\T({\lambda,\omega}) $, of a regular tree
graph $ \T{} $ with branching number $ K \geq 2 $
the AC spectrum  of   $ -\Delta_{\T({\lambda,\omega})}$ is continuous
at $\lambda=0$    in
the sense that for any interval $ I \subset \R $ and almost all $ \omega $:
     \begin{equation}
		\lim_{\lambda \to 0 } \; \mathcal{L}\big( I \cap
\sigma_{\rm ac}(-\Delta_{\T({\lambda,\omega})})\big)
     = \mathcal{L}\big(I \cap \sigma_{\rm ac}(-\Delta_{\T{}}) \big)
\end{equation}
where $\mathcal{L}(\cdot)$ denotes the Lebesgue measure.
\end{theorem}
\begin{remarks}
\begin{nummer}
\item  As is generally known by ergodicity arguments \cite{CaLa90,PF,AcKl92}, 
and in our case also by the $0$-$1$ law
for the sigma-algebra of events measurable at infinity, which is applicable
through  Theorem~\ref{thm:spec}, for almost all $ \omega $ the AC spectrum
of $-\Delta_{\T({\lambda,\omega})}$ is given by a certain non-random set.
\item The assumption on the
distribution of  $ \left\{ \omega_e \right\}_{e \in \mathcal{E}} $
can be relaxed:  the present proof readily extends to the class of
random graphs  where the distribution of these variables is
{\em stationary} under the endomorphisms of the tree $ \T $ 
and {\em weakly correlated} in the sense of \cite[Def.~1.1]{ASW05}.
     \item To better appreciate the continuity asserted in
Theorem~\ref{thm:main},
  one may note that the analogous statement is not expected to be true
  in case the disorder is restricted to be radially symmetric, i.e.,
  $\omega_e =  a_{{\rm dist}\{e,0\}}$ with $\{a_n\}$  a collection of iid
  random variables.   In this case, the AC spectrum coincides with that
  of a one-dimensional Sturm-Liouville operator. In view of related results 
about Anderson localization in one dimension \cite{CaLa90,PF,Min1,Min2} one may expect 
(though we are not aware of a published proof) that also here
 localization sets in at any non-zero
  level of disorder.
\end{nummer}
\end{remarks}

\section{An outline of the argument}
A generally useful tool for the study of the  spectral and
dynamical properties of any quantum graph is provided by the Green
function.   For tree graphs, we find it particularly useful
to consider a related quantity, which is an extension of the
Weyl-Titchmarsh function familiar from the context of
Sturm-Liouville or Schr\"odinger operators on a line.
Before outlining the main steps in the derivation of
Theorem~\ref{thm:main}, we shall introduce this function
and its key properties, first somewhat informally
through its appearance in a scattering problem.

\subsection{A scattering perspective}
As noted by Miller and Derrida~\cite{MilDer93}, one may obtain a scattering
perspective on extended states by considering a setup in which a wire
$ \mathbb{W}_x $
is attached to a tree graph $ \T $ at an interior point $x$ of an
edge. Particles
of energy $E$ and decay rate $\eta$ are sent at a steady rate down
this wire.  In the corresponding steady state, the quantum
amplitude $ \psi $ for observing a particle at a point is given by
a function satisfying
$(-\Delta_{\T\cup\mathbb{W}_x} -z ) \,\psi = 0$, were $z=E+i\eta \, $
and $- \Delta_{\T\cup\mathbb{W}_x}\, $ is a self adjoint Laplacian on the
union of the graph and the wire,  defined
with suitable BC for the  {\em three} segments
meeting at the point of contact.  For the latter,
we assume here that it will be appropriate to take the Kirchhoff conditions.

As follows from Theorem~\ref{prop:unique} below, on  the
two subgraphs $ \T^+_x $ and $ \T^-_x $,  produced by cutting
$\T$  at $x$, the
above differential equation has a unique  -- up to a multiplicative constant --
square-integrable solution $\psi^+$ and correspondingly $\psi^-$.
Thus $ \psi$  takes the form:
\begin{equation}
\psi(y;z) \ = \ \left\{ \begin{array}{c@{$\quad$}l}
            e^{i \sqrt{z} (y-x)} + r(x;z)\, e^{- i \sqrt{z} (y-x)} &
\mbox{along the
wire} \\[1ex]
            \psi^{\pm}(y;z) & \mbox{along the graph}
            \end{array}
                    \right.
\end{equation}
where $r(x;z) $ is  the reflection coefficient,
and the three branches
are linked through the Kirchhoff conditions:
\begin{align}
& \psi^+(x;z) \  =\ \psi^-(x;z) \  =\ 1 + r(x;z) \nonumber \\
& \frac{\partial}{\partial x} \psi^{+} (x;z) \   - \frac{\partial}{\partial x} \psi^{-}(x;z) \  = \
i \sqrt{z} \; \big( 1-r(x;z) \big)
\end{align}
with the differentiation taken in the direction
away from the root of $ \T $.
The above relations yield
\begin{equation}  \label{eq:reflection}
i\sqrt{z}\;\, \frac{1-r(x;z)}{1+r(x;z)} \ = \ R^+(x;z) \ +\  R^-(x;z)
\end{equation}
where $ R^\pm = \pm \left(\partial \psi^{\pm}/ \partial x\right) / \psi^\pm  $.

  From the scattering perspective the graph absorbs some of
the current directed at it, i.e., conducts it to infinity,
if and only if $|r(x;z)|<1$. A simple consequence of
\eqref{eq:reflection} is the equivalence
\begin{equation}
|r(x;E)|<1  \qquad   \Leftrightarrow \qquad \Im \left( R^+(x;E) \ +\
R^-(x;E) \right) > 0  \, .
\end{equation}

As it turns out  $R $ also plays a direct role in the
spectral theory of $-\Delta_{\T}$:  the diagonal of its
Green function  is given by
\begin{equation}\label{eq:Green}
G_{\T}(x,x;z) \ = \   - \left( R^+(x;z) \ +\  R^-(x;z) \right)^{-1}  \; .
\end{equation}
By the theorem of  de la Vall\'ee Poussin, the AC component of the spectral measure, associated with the
function in \eqref{eq:Green}, is  $\pi^{-1} \Im G_{\T}(x,x;E+i0)\, dE$.  Therefore,
there is a relation between the occurrence of the AC spectrum, the
ability of the graph to conduct
current to infinity, and the non-vanishing of $\Im R^\pm(x;E)$.\\
Let us note that the reflection coefficient for the version of the above
experiment in which the particles are sent towards only the forward
subtree $\T^+_x$, is given by a version of \eqref{eq:reflection} with
only $R^+(x;z)$ on the right side, and similarly for  $\T^-_x$.

\subsection{Tree extension of the Weyl-Titchmarsh function}\label{Sec:WT}

We shall now follow the somewhat informal introduction above with a more
careful definition of the functions $R^\pm$.
For this purpose the following statement plays an important role.
\begin{theorem} \label{prop:unique}
Let $\G$ be a connected metric graph with a selected ``open'' vertex $u$ which
has exactly one adjacent edge.
Let $-\Delta_{\G,u}$ be the symmetric Laplacian defined 
with self-adjoint BC on all vertices excepting the
open vertex, where it is required that both $\psi(u)=0$ and $\psi'(u)=0$.
Then:
\begin{indentnummer}
\item
For any $z \in \C^+ := \{ z \in \C \, : \, \Im z > 0 \} $, the space of square-integrable
solutions of $(-\Delta_{\G,u}^* - z) \, \psi =0$, with $-\Delta_{\G,u}^*$
the adjoint operator, is one dimensional.
\item
The solution $\psi(x;z)$  and its derivative $\psi'(x;z)$
do not vanish on any point which disconnects $\G$.
\item  Normalized so that  $\psi(u;z)=1$, both $\psi(x;z)$ and $\psi'(x;z)$
are analytic for $z \in \C^+ $ and all $ x \in \G $.
\end{indentnummer}
\end{theorem}
We note that $- \Delta_{\G,u} $ is \emph{not} self-adjoint. 
The proof of this theorem is given in Appendix~\ref{app:more}.\\

The following corollary is a relevant implication for trees. Throughout, we denote by $\psi^{\pm }(x;z|u)$
the functions described in Theorem~\ref{prop:unique}
which correspond to the two subtrees, $\T^{\pm}_u$,
into which $\T$ is split at $u$, with $u$ serving as the open vertex. 
We fix their normalization such that $ \psi^{\pm }(u;z|u) = 1 $.
\begin{corollary}\label{cor:R}
Along the edges of a metric tree $\T$, the  ratio
\begin{equation}\label{def:R}
R^\pm(x;z) := \pm \;
\frac{1}{\psi^{\pm } (x;z|u)} \,
\frac{\partial }{\partial x}{\psi^{\pm } (x;z|u)}  \,
\end{equation}
does not depend on $u$ as long as $x$ stays in $\T^{\pm}_u$.
\end{corollary}

\begin{definition}\label{def:WT}
We shall refer to the above $R^\pm$  as the
\emph{(generalized) Weyl-Titchmarsh (WT) functions}.
\end{definition}

These functions have a number of properties which
are used in the proof of our main result.  If not obvious,
their derivation is given in Appendix~\ref{app:more}.
\begin{enumerate}
\item({\em Relation with the Green function\/}) The generalized WT
function may be related to the diagonal elements
of the Green function which is defined on  $\T^+_x$, with the
$\alpha\neq 0$ BC at $x$, as
\begin{equation}\label{eq:RGreen}
	R^+(x;z) \ =\  \cot\alpha - \frac{1}{G^\alpha_{\T_x^+}(x,x;z)}  \, ,
\end{equation}
and similarly for $R^-$.
\item({\em Boundary values\/})  The  function has the
Herglotz-Nevanlinna property \cite{Duren}:
it is analytic for $z\in \C^+$ with $\Im R^\pm(x;z) >  0$ when
$\Im z >0$.   By a standard implication, for
each $x$ the limit
\begin{equation}
R^\pm(x;E+i0) := \lim_{\eta \downarrow 0} R^\pm(x;E+i\eta)
\end{equation}
exists for Lebesgue almost every $E\in \R$.
\item({\em Evolution along the tree\/})
  The values $ R^+_e(\cdot;z) $ at two opposite ends of an
edge $ e $ are related by a M\"obius transformation, which integrates
the Riccati equation:
\begin{equation}\label{eq:Riccati}
\frac{\partial }{\partial x} R^+(x;z) + z + R^+(x;z)^2 = 0 \, .
\end{equation}
Over each vertex $ R^+(\cdot;z) $ is additive thanks to \eqref{eq:Kirch}:
\begin{equation}\label{eq:add}
             R^+_e\big(L_e;z\big)   \ = \
\sum_{f \in \mathcal{N}_e^+} R^+_{f}\big(0;z\big) \, .
\end{equation}
\item({\em Relation with the current\/})  For each $u$, the
quantity
\begin{align}
J^+(x,z|u) & :=   \Im \left[ \overline{\psi^+(x;z|u)} \, \frac{\partial}{\partial x}
\psi^+(x;z|u)
  \right]   \notag \\
	   & = |\psi^+(x;z|u)|^2\, \Im R^+(x;z)
 \geq 0 \label{eq:defcurrent1}
\end{align}
represents a current. It is additive at the vertices and conserved along
the edges for real $z$.  For $z\in \C^+$
the current is decreasing in the direction away from
the root:
\begin{equation}\label{eq:defcurrent}
\frac{\partial}{\partial x} J^+(x;z|u) =  - |\psi^+(x;z|u)|^2 \, \Im
z \ \leq \  0 \, .
\end{equation}
At interior vertices the net current flux is zero.
\end{enumerate}
\subsection{The core of the argument}
\label{subsec:outline}
We now have the requisite tools to outline the
proof of the persistence of the AC spectrum
under weak disorder.
A key element in our analysis is to show  that for small
$(\lambda,\eta) $, the WT function
$R^+(x;E+i\eta,\lambda,\omega)$
does not depend  much on $ \omega $. At each point its distribution is narrowly
peaked around a value which may only depend on $(\lambda,\eta)$,
and the relative location of the point within the edge.
By the rules of the evolution of $R^+ $
along an edge, which are described above, it follows that
for  $(\lambda,\eta) \to (0,0)$ the limit of the ``typical''
value of $R^+_e\big(0;z,\lambda,\omega \big)$, or
more precisely any  accumulation point of such, obeys a
M\"obius evolution whose unique periodic solution is given by the
WT function of the regular tree $\T{} $.  The continuity
then readily follows, though some care is needed in
the presentation of the argument.  In this part, we
employ the strategy which was presented in~\cite{ASW05}.

It should be appreciated that the asymptotic lack of
dependence of $R^+(x,z; \lambda,\omega) $ on
$ \omega $ is not just a trivial consequence of the smallness of
$\lambda$ since this parameter affects an infinite number
of random terms.  As commented above, it is natural to expect the
corresponding  statement to fail when the  disorder is radial,
with $\omega $ given by radially symmetric but otherwise iid 
random variables.   To streamline the notation, in various 
places the dependence of $ \psi^+ $ and $R^+ $
on $\lambda$ and $\omega$ will be suppressed.  \\ 

The first statement establishing a reduction of fluctuations concerns
$\Im R^+(x;z)$.
For that the starting point is \eqref{eq:defcurrent1} by which
$|\psi^+(x;z|0)|^2\cdot \Im R^+(x;z) $
gives  the flux at $x$ of a conserved current.  The current is
injected at the root and at each vertex it is split
among the forward directions.  It is significant that the first factor
takes a common value among  the different
forward directions,   
the second factor is independently distributed, and, furthermore,
it has the same distribution  as the total current $\Im R^+(0;z)$.
It follows that
      \begin{equation}\label{eq:recur}
	\frac{\frac{1}{K} \sum_{e \in \mathcal{N}_0^+} \Im
	R^+_{e}\big(0;z,\lambda,\omega\big) }{\Im
R^+_0\big(0;z,\lambda,\omega\big)}
	\leq \frac{\left|\psi^+_0\big(0;z, \lambda,\omega |0\big) \right|^2}{K\,
		\big|\psi^+_f\big(0; z, \lambda,\omega 
		|0\big) \big|^2 }  \, .
      \end{equation}
This expresses  current conservation/attrition, and
for $ \Im z = 0 $ holds as equality.
Here $ f  \in \mathcal{N}_0^+ $ is an arbitrary edge forward to that of the root, 
and due to the particular normalization
chosen (before Corollary~\ref{cor:R}) the numerator on the right side is actually one.   
Our argument  proceeds by combining two essential observations:  
\begin{enumerate}
\item
By the Jensen inequality  the expectation value
of the logarithm of
the left side of \eqref{eq:recur}
 is non-negative.  The inequality can be strengthened to show
 that the above expectation value provides an upper bound on
  a positive quantity which expresses the
relative width of the distribution of
$ \Im  R^+_0\big(0;z,\lambda,\omega\big)$.
\item  The expectation of  the logarithm of
the right side of \eqref{eq:recur} is a quantity which it is natural to
regard as a Lyapunov exponent,
	\begin{equation}\label{eq:defgamma}
	\gamma_\lambda(z) := -
	\mathbb{E}\left[ \log \sqrt{K} \;
	\frac{\big|\psi^+_f\big(0; z, \lambda, \cdot |0\big) \big|}{\big| 
	\psi^+_0\big(0;z, \lambda, \cdot |0\big) \big|} \right]  \, ,  
	\end{equation}
For $\lambda=0$, this Lyapunov exponent   vanishes  for almost every
$ z \in \sigma_{\rm ac}(-\Delta_{\T{}})$.  Furthermore, the
average  of $\gamma_\lambda(E+i\eta)$ over any energy  interval is
continuous in $(\lambda,\eta)$.
\end{enumerate}

The above mentioned  improvement of the Jensen inequality is
summarized in the following statement, which is a
consequence of \cite[Lemma~3.1 and Lemma~D.2]{ASW05}.

\begin{lemma}\label{lemma:jensen}
Let $ \{ X_j \}_{j=1}^K $ be a collection of $K\ge 2$  iid
positive random variables, and $X$ a variable of the same
distribution. Then for any $a\in (0,1/2]$:
    \begin{equation} \label{jensen}
\mathbb{E}\left[  \log \left( \frac{1}{K} \sum_{j=1}^K X_j \right)
    \right] \  \geq \   \mathbb{E}\left[ \log X \right] \
                        + \  \frac{a^2}{4}
    \delta \left( X, a \right)^2 \, .
    \end{equation}
    where $ \delta(X,a) $ is the relative $a$-width of
    $ X$, which is defined below.
    \end{lemma}
\begin{definition}
The relative $a$-width of the  distribution of
a positive random variable $X$, at  $a\in (0,1/2]$, is
    \begin{equation}
       \delta(X,a) := 1 - \frac{\xi_-(X,a)}{\xi_+(X, a)}
                \end{equation}
with $ \xi_-(X,a) = \sup \{\, \xi \, :\,
\mathbb{P}(X < \xi) \le a\} $ and $ \xi_+ (X, a) =
\inf \{\,  \xi \, :\,  \, \mathbb{P}(X>\xi) \le a\} $.
\end{definition}
A number of  useful rules of estimates of the the relative width
of a distribution are compiled in \cite[Appendix~D]{ASW05}.

We shall now turn to the two key properties of the
   Lyapunov exponent which were mentioned above.


\section{A Lyapunov exponent and its continuity}\label{subsec:Lya}

We shall refer to $ \gamma_\lambda(z) $ which is defined by
\eqref{eq:defgamma}
as the Lyapunov exponent of the randomly deformed tree~$
\T({\lambda,\omega}) $.
The following theorem collects some of its properties.
Of particular relevance is that the integral of $ \gamma_\lambda(E+i \eta) $
over $ E \in \sigma_{\rm ac}(\T) $ is small for small
$ \lambda $ and $ \eta $.
	\begin{theorem}\label{thm:Lyaco}
The Lyapunov exponent  $\gamma_\lambda(z) $ has the following properties:
\begin{indentnummer}
\item  As a function of $z \in \C^+$, it  is positive and harmonic with
	$ \gamma_{\lambda}(i \eta)/ \eta \to 0 $ for $\eta \to \infty $.
\item
For $ \lambda = 0 $, it vanishes on the AC spectrum: $ \gamma_0(E+i0) = 0 $
for Lebesgue-almost all $E\in \sigma_{\rm ac}(-\Delta_{\T{}})$.
\item For any $z\in \C^+$, $\gamma_\lambda(z+i\eta)$ is jointly continuous
in $(\lambda,\eta) \in \R\times [0,\infty) $.
	\item For any
         $  [a,b] \subset \sigma_{\rm ac}(-\Delta_{\T{}}) $:
         \begin{equation}\label{eq:cont}
           \lim_{\substack{\lambda \to 0 \\ \eta \downarrow 0}}\;
\int_a^b \gamma_\lambda(E+i\eta) \, d E  =  0 .
         \end{equation}
\end{indentnummer}
\end{theorem}
\begin{proof}
\begin{nummer}
\item From \eqref{eq:defgamma} and \eqref{def:R} it follows that
$ \gamma_\lambda(z) $ is the negative of the real part of the
Herglotz-Nevanlinna function
\begin{equation}\label{eq:w}
	 w_\lambda(z) :=  \log \sqrt{K} + \mathbb{E}\left[
	\int_0^{L_0(\lambda)}\!\! R^+_0\big(l;z,\lambda\big) \, dl \right],
\end{equation}
and hence it is harmonic.
The positivity of $ \gamma_\lambda(z) $ follows from \eqref{eq:defcurrent1} 
and the Jensen inequality, which yield
\begin{equation}\label{eq:currentLya}
2 \gamma_\lambda(z) \geq \mathbb{E}\left[ \log \frac{J^+_0(0;z|0)}{J^+_0(L_0(\lambda,\cdot);z|0)} \right] > 0
\end{equation}
due to the
current loss \eqref{eq:defcurrent} on every edge for $ z \in \mathbb{C}^+ $.
The statement of asymptotics derives from \eqref{eq:boundsol} and 
the bound \eqref{eq:boundR} in Appendix~\ref{app:more}.
\item The vanishing of $\gamma_0$ along $\sigma_{\rm ac}(-\Delta_{\T})$
is a consequence of the $ \Im z \downarrow 0 $
limit of~ \eqref{eq:recur} and the fact that
$ R^+_e(0;z,0) $ is independent of $ e $, with $0 < \Im
  R^+_e(0;E+i0,0) < \infty $ for Lebesgue-almost all
$ E \in \sigma_{\rm ac}(-\Delta_{\T{}})$.
\item From~\eqref{eq:defgamma} and \eqref{eq:explsol} together with
the dominated convergence theorem, which is applicable 
due to \eqref{eq:boundsol} and Theorem~\ref{thm:profR},
we conclude that the continuity of $ \gamma_\lambda(z+i\eta) $ follows 
from that of $ R_0(0;z+i\eta,\lambda,\omega) $. The latter
is derived using the argument in the proof of Theorem~\ref{thm:profR}(iv).
\item By virtue of~(ii)
it suffices to prove that
\begin{equation} \label{eq:convw}
\lim_{\lambda, \eta \to 0 }\;
\int_a^b \gamma_\lambda(E+i\eta) \, d E = \int_a^b \gamma_0(E+i0) \, d E.
\end{equation}
To do so, we note that the integrals in \eqref{eq:convw} can be
associated with the (unique)
Borel measure $ \sigma_{(\lambda,\eta)} $ corresponding to the
positive  harmonic function $ h_{(\lambda,\eta)}(z) =
\gamma_{\lambda}(z + i \eta) $ (cf.\ \eqref{eq:repharm} below). Since $ w_\lambda( \cdot + i \eta)  $ 
has the Herglotz-Nevanlinna
property, the harmonic
conjugate of  $ h_{(\lambda,\eta)} = - \Re w_\lambda( \cdot + i \eta) $ 
has a definite sign and hence locally integrable boundary values
\cite[Thm.~1.1]{Duren}. Therefore, the measure
$ \sigma_{(\lambda,\eta)} $ is purely AC
\cite[Thm~3.1 \& Corollary~1]{Duren}  for all $ (\lambda,\eta) \in
[0,1]^2 $ and given by
\begin{equation}
\sigma_{(\lambda,\eta)}\left[a,b\right] = \int_a^b
\gamma_{\lambda}(E + i \eta) \, dE \, .
\end{equation}
The assertion thus follows from~(iii) and Lemma~\ref{lemma:harm} below.
\qed
\end{nummer}
\end{proof}
The last part of the preceding proof was based on the following
general convergence result for sequences of harmonic functions.
Recall (cf.\ \cite{Duren,Kot85}) that every positive harmonic
function $ h: \C^+ \to (0,\infty) $
which satisfies $ \lim_{\eta \to \infty} h(i\eta)/\eta = 0 $
admits the representation
\begin{equation}\label{eq:repharm}
	h(z) =  \int_\R \frac{\Im z}{|E-z|^2} \; \sigma(dE)
\end{equation}
with some positive Borel measure $ \sigma $ on $ \R $ with $
\int_{\R} (E^2 + 1)^{-1} \sigma(dE) < \infty $.
\begin{lemma}\label{lemma:harm}
	Let $ h_n$, $h : \C^+ \to (0,\infty) $
         be positive harmonic functions with
	$ \lim_{\eta \to \infty} $\hspace{0pt}$h_{n}(i\eta)/\eta = 0 $ and similarly for $ h $.
	Suppose that for all $ z \in \C^+ $
	\begin{equation}\label{eq:ptwcon}
	\lim_{n \to \infty} h_n(z) = h(z).
	\end{equation}
	Then their associated Borel measures converge vaguely, $
\lim_{n \to \infty} \sigma_n = \sigma $.
\end{lemma}
The proof is an immediate consequence of the representation \eqref{eq:repharm} 
and \cite[Prop.~4.1]{HLMW01} (see also \cite[Lemma~5.22]{PF}).

\section{Fluctuation bounds}
Proceeding along the lines outlined in
    Subsection~\ref{subsec:outline},
    we shall now show that a small Lyapunov exponent $
\gamma_\lambda(z) $ implies the sharpness of the distribution of both the
imaginary part and the modulus of a certain
linear function
of $ R^+_0(0;z,{\lambda,\omega}) $.
\begin{theorem}\label{thm:imfoc}
For any $ \lambda \in
\R $, $ z \in \C^+ $ and $ a \in (0,1/2] $:
          \begin{align}\label{eq:imfoc1}
                 & \delta\big(\Im R^+_0(0;z,\lambda,\cdot),a \big)^2    \leq
                \frac{8}{ a^2} \, \gamma_\lambda(z), \\[1.5ex]
                 & \delta\Big( \Big| \cos\big(\sqrt{z} L_0(\lambda,\cdot)\big) +
\frac{\sin\big(\sqrt{z} L_0(\lambda,\cdot)\big)}{\sqrt{z}} R^+_0(0;z,\lambda,\cdot)
            \Big|^2 ,a \Big)^2 \notag \\	
		& \mkern175mu
            \,      \leq \,  512 \; \frac{(K+1)^2}{a^2} \, \gamma_\lambda(z) \, . \label{eq:imfoc2}
          \end{align}
\end{theorem}
\begin{proof}
The derivation of~\eqref{eq:imfoc1} starts from
the relation
\begin{equation}\label{eq:stopfl1}
2 \gamma_\lambda(z) \geq \mathbb{E}\Big[ \log \Big(
\frac{1}{K}  \sum_{f \in \mathcal{N}_0^+}
                    \Im R^+_{f}\big(0;z,\lambda,\cdot\big) \Big) \Big]
            - \mathbb{E}\left[ \log\left( \Im
R^+_0\big(0;z,\lambda,\cdot\big) \right)\right]
\end{equation}
which is obtained by taking the expectation of the logarithm of
\eqref{eq:recur}.  Applying the improved Jensen
inequality~\eqref{jensen}, and using the fact that
$  \Im R^+_f{\big(0;z,\lambda,\omega}\big) $  are iid for
$ f \in \mathcal{N}_0^+$, the right side of
\eqref{eq:stopfl1} is bounded from below
by $a^2 \, \delta\left(\Im R^+_0(0;z,\lambda,\cdot),a\right)^2   /4 $.
This implies \eqref{eq:imfoc1}.\\
The proof of \eqref{eq:imfoc2} starts by observing that the quantity
in its left side can be identified with the right side of \eqref{eq:recur}:
\begin{align}
& \cos\big(\sqrt{z} L_0(\lambda,\omega)\big) +
\frac{\sin\big(\sqrt{z} L_0(\lambda,\omega)\big)}{\sqrt{z}} \,
R^+_0(0;z,\lambda,\omega) \notag \\
& = \psi^+_0\big(L_0(\lambda,\omega); z,\lambda,\omega |0\big) \, .
\end{align}
This follows from \eqref{eq:explsol} in Appendix~\ref{app:more}.
Setting $ X := J^+_0(L_0(\lambda,\cdot);z|0)/ J^+_0(0;z|0)$ and using the definition of the current,
the left side in \eqref{eq:imfoc2} therefore equals
\begin{align}\label{eq:R12}
& \delta\Bigg(\frac{\Im R^+_0\big(0;z,\lambda,\cdot\big)}{\sum_{f
\in \mathcal{N}_0^+} \Im R^+_{f}\big(0;z,\lambda,\cdot\big) } X , a\Bigg)
\notag \\
& \leq \delta\Bigg(\frac{\Im R^+_0\big(0;z,\lambda,\cdot\big)}{\sum_{f
\in \mathcal{N}_0^+} \Im R^+_{f}\big(0;z,\lambda,\cdot\big) }, \frac{a}{2}\Bigg)
	+ \delta\Bigg(X, \frac{a}{2}\Bigg) \, ,
\end{align}
where the inequality results from the additivity of the relative width under multiplication \cite[Lemma~D.1]{ASW05}.
This additivity and the invariance under inversion \cite[Lemma~D.1]{ASW05} ensures that
the first term on the right side of \eqref{eq:R12} is bounded from
above by
\begin{align}
& \delta\Big(\Im R^+_0\big(0;z,\lambda\big), \frac{a}{2 (K+1)} \Big) +
\delta\Big(\sum_{f
\in \mathcal{N}_0^+} \Im R^+_{f}\big(0;z,\lambda\big), \frac{a K}{2 (K+1)}
\Big)  \notag \\
& \leq 2 \;
\delta\Big(\Im R^+_0\big(0;z,\lambda\big), \frac{a}{2 (K+1)} \Big) \leq \frac{8 \sqrt{2} \,(K+1)}{a} \sqrt{ \gamma_\lambda(z)} \, .
\end{align}
Here the first inequality results from the rules of addition of
iid random variables \cite[Lemma~D.1]{ASW05}. The second one is a consequence of (\ref{eq:imfoc1}).
The second term in on the right side of \eqref{eq:R12} is bounded from above according to
$  \delta(X,a/2) \leq 2 \sqrt{\gamma_\lambda(z)/a } $.
This follows from \eqref{eq:currentLya} and the simple bound
\begin{equation}
	\delta(X,a)^2 \leq (1 -\xi_-(X,a))^2 \leq - \ln \xi_-(X,a) \leq - \frac{\mathbb{E}\left[ \ln X \right]}{a} ,
\end{equation}
valid for all random variables $ X $ taking values in $ (0,1] $.
Combining the above estimates, we arrive at \eqref{eq:imfoc2}.
\qed
\end{proof}

\section{Stability of the Weyl-Titchmarsh function under weak disorder}
\subsection{The main stability result}

Our goal in this section is to show that the boundary values  of the WT function
are continuous at $ \lambda = 0 $ in a certain
distributional sense as long as  $ E \in \sigma_{\rm
ac}(-\Delta_{\T{}}) $. Here
the distribution refers to the joint dependence on the energy
and the randomness.  The result to be derived is:
\begin{theorem}\label{thm:stoch}
Let $ I \subset \sigma_{\rm ac}(-\Delta_{\T{}}) $ be an interval.
Then the WT function converges
in $ \mathcal{L}_I \otimes \mathbb{P} $-measure, i.e., for all $ \varepsilon > 0 $:
\begin{equation}\label{eq:stoch}
\lim_{\lambda \to 0} \mathcal{L}_I \otimes \mathbb{P} \left\{
            \big| R^+_0(0;E+i0,\lambda,\omega) - R^+_0(0;E+i0,0) \big| > \varepsilon \right\} = 0
\end{equation}
where $ \mathcal{L}_I $ denotes the
Lebesgue measure on $ I $.
\end{theorem}

The above statement will be derived in this section
by proving, in
Theorem~\ref{thm:meas} which appears below, that
for all $ \varepsilon > 0 $ and all sequences $ (\lambda,
\eta) $ converging to zero
\begin{equation}\label{eq:disconv}
\lim_{\lambda,\eta\to 0 }
\, \mathcal{L}_I \otimes \mathbb{P} \left\{
            \big| R^+_0(0;E+i\eta,\lambda,\omega) - R^+_0(0;E+i0,0) \big| >
\varepsilon \right\} = 0  \, .
\end{equation}

Before we delve into the proof of the statements which lead to
Theorem~\ref{thm:stoch} let us note that  it implies our main
claim.
\begin{proof}[of Theorem~\ref{thm:main}; assuming Thm.~\ref{thm:stoch} ]
Since $  \sigma_{\rm
ac}(-\Delta_{\T(\lambda,\omega)}) $ coincides almost-surely
with a non-random set, it suffices to show that 
\begin{equation}
 \lim_{\lambda \to 0} \; \mathbb{E}\left[\mathcal{L}\big( I \cap \sigma_{\rm
ac}(-\Delta_{\T(\lambda,\cdot)}) \big)\right] = \mathcal{L}\big(I\big) \, .
\end{equation}
We start the proof of this relation by observing that
\begin{align}
\mathcal{L}\big(I\big) & \geq
\mathbb{E}\left[\mathcal{L}\big( I \cap \sigma_{\rm
ac}(-\Delta_{\T(\lambda,\cdot)}) \big)\right] \notag \\
& \geq \mathcal{L}_I\otimes \mathbb{P}\,\big\{
            0 < \Im R^+_0(0; E+i0,\lambda,\omega) < \infty \big\}
\label{eq:lower1}
\end{align}
where the second inequality is due to Theorem~\ref{thm:spec}. 

For
any $ \varepsilon > 0 $ the set on the right side includes the collection of
$ (E,\omega)  $ for which $
\varepsilon < \Im R^+_0(0; E+i0,0) < \infty $ and
$ \big| \Im R^+_0(0; E+i0,\lambda,\omega) - \Im R^+_0(0; E+i0,0) \big| \leq
\varepsilon $. Accordingly, the right side of~\eqref{eq:lower1}
is bounded below by the difference of
\begin{equation}\label{eq:meas1}
            \mathcal{L}_I\otimes \mathbb{P}\,\big\{
            \big| \Im R^+_0(0; E+i0,\lambda,\omega) - \Im R^+_0(0; E+i0,0) \big|
            \leq \varepsilon \big\}
\end{equation}
and
\begin{equation}\label{eq:meas2}
\mathcal{L}_I\big\{ \Im R^+_0(0; E+i0,0) \in [0,
\varepsilon] \cup \{\infty \} \big\}.
\end{equation}
As $ \lambda \to 0 $ the measure in \eqref{eq:meas1} converges to $
\mathcal{L}(I) $ by Theorem~\ref{thm:stoch}. Moreover,
as $ \varepsilon \downarrow 0 $ the measure in \eqref{eq:meas2}
converges to zero.
\qed
\end{proof}

\subsection{Convergence in measure}

In order to derive Theorem~\ref{thm:stoch}, we shall consider the
distribution under the measure  $ \mathcal{L}_I \otimes \P $ of
the joint values of $ E $, $ \left\{R^+_e(0;E+i\eta,{\lambda,\omega}) \right\}_{e \in \mathcal{E}} $, and 
$  \left\{ L_e({\lambda,\omega}) \right\}_{e \in \mathcal{E}} $.
In the following, $L_{\max}$ stands for some uniform upper bound on
$ L_e(\lambda,\omega)  $, which exists due to the boundedness of the
random variables.
The setup is similar to that employed in \cite{ASW05}.

\begin{definition}
Let $ (\lambda, \eta) \in [0,1]^2 $ and $ I \subset \sigma_{\rm
ac}(-\Delta_{\T{}}) $. The
Borel measure $ \nu_{(\lambda,\eta)} $ on $ I \times \C^\mathcal{E}
\times [0,L_{\max}]^\mathcal{E} $ is the measure
induced  by $ \mathcal{L}_I \otimes \P $
under the mapping
\begin{equation}\label{def:mun}
            (E,\omega) \mapsto \left(E,
\left\{R^+_e\big(0;E+i\eta,{\lambda,\omega}\big) \right\}_{e \in \mathcal{E}} ,
            \left\{ L_e({\lambda,\omega}) \right\}_{e \in
\mathcal{E}} \right).
\end{equation}
Moreover, its
            $ E $-conditional distribution on $ \C^\mathcal{E} \times
[0,L_{\max}]^\mathcal{E} $
is abbreviated by $\nu_{(\lambda,\eta)}^E $.
\end{definition}
\begin{remarks}
\begin{nummer}
\item The above definition relies on the fact that one may identify
the edge sets $ \mathcal{E} $ of $ \T(\lambda,\omega) $ corresponding to
different values of $ \lambda $ and/or $ \omega $.
\item In case $ (\lambda,\eta) = (0,0) $ the measure $ \nu_{(\lambda,\eta)} $
is a product of the Lebesgue measure and products of Dirac measures:
\begin{equation}
            \nu_{(0,0)} = dE \, \bigotimes_{e\in \mathcal{E}}
\delta_{R^+_0(0;E+i0,0)} \,
            \bigotimes_{e\in \mathcal{E}} \delta_{L}.
\end{equation}
\item The family of finite measures $ \nu_{(\lambda,\eta)} $
is tight. Indeed,
the bound \eqref{eq:boundR} in Appendix~\ref{app:more} and 
arguments as in \cite[Prop.~B.1 \& Lemma~B.1]{ASW05} show that
\begin{equation}
            \inf_{t > 0} \; \sup_{(\lambda,\eta)\in [0,1]^2} \;
\nu_{(\lambda,\eta)}\left( \big| R_e \big| > t \right) = 0
\end{equation}
for all $ e \in \mathcal{E} $. Accordingly, every sequence of
measures $ \nu_{(\lambda,\eta)} $ corresponding
to $ (\lambda, \eta) \to (0,0) $ has weak accumulation points.
\end{nummer}
\end{remarks}
The issue now is to show that all of the above mentioned accumulation
points coincide.
\begin{theorem}\label{thm:meas}
In the sense of weak convergence:
\begin{equation}\label{eq:weakconv}
\lim_{\lambda,\eta\to 0 } \;
\nu_{(\lambda,\eta)} = \nu_{(0,0)}.
\end{equation}
\end{theorem}
The proof of this theorem closely follows ideas in \cite{ASW05}, and rests
on the following two lemmas.

We first show that all accumulation points of the sequence in
\eqref{eq:weakconv} are supported on points
satisfying the limiting recursion relation.
\begin{lemma}\label{lem:twop}
Let  $ \nu $ be a  (weak) accumulation point for the family of measures
$\nu_{(\lambda, \eta)}$, with the parameters $ (\lambda,
\eta) $ in $ [0,1]
\times (0,1] $ converging to $(0,0)$.
Then
\begin{indentnummer}
\item the limiting recursion relation
                 \begin{multline} \label{eq:reclim}
                    \Big(\cos\big(\sqrt{E} L_e\big)  +
\frac{\sin\big(\sqrt{E} L_e\big)}{\sqrt{E}}  R_e \Big)
            \sum_{f \in \mathcal{N}^+_e} R_{f} \\
                    =   \cos\big(\sqrt{E} L_e\big) R_e - \sqrt{E}
\sin\big(\sqrt{E} L_e \big) \mkern50mu
                  \end{multline}
holds for $ \nu $-almost all $ (E,R,L) \in I\times
\C^\mathcal{E}\times [0,L_{\max}]^\mathcal{E}  $.
\item  the lengths are $ \nu
$-almost surely constant, $ L_e = L $ for all $ e \in \mathcal{E} $.
\item  the variables $ \{ R_e \}_{e \in
\mathcal{E}} $ are identically distributed $ \nu $-almost surely.
\item  for Lebesgue-almost all $ E \in I $
                    there exist $ \mathcal{I} \in [0,\infty) $ and $ \mathcal{M} \in
[0,\infty) $ such that for all $ e\in \mathcal{E} $
\begin{equation}\label{eq:ImMod}
              \Im R_e = \mathcal{I} \quad \mbox{and} \quad \Big|
\cos\big(\sqrt{E} L\big) + \frac{\sin\big(\sqrt{E}
L\big)}{\sqrt{E}}
            R_e         \Big|
            =  \mathcal{M}
\end{equation}
$ \nu^E $-almost surely.
           \end{indentnummer}
\end{lemma}
\begin{proof}
\begin{nummer}
\item The fact that the accumulation points obey the limiting
recursion relation which is the $(\lambda,\eta)=(0,0)$ version
of \eqref{eq:Moebius} and \eqref{eq:add}, is a
consequence of  the general principle proven in
\cite[Prop.~4.1]{ASW05}.
\item This statement is implied by the pointwise
convergence $
\lim_{\lambda \to 0} L_e(\lambda,\omega) = L $.
\item The
claim follows from the fact that all prelimit quantities
 are identically distributed.
\item We fix $ e \in \mathcal{E} $. Then
Theorem~\ref{thm:imfoc} and Theorem~\ref{thm:Lyaco} yield
\begin{align}
            & \lim_{\lambda,\eta\to 0 }\,\int_I \delta\big(\Im R^+_e(0;E+i\eta,\lambda,\cdot),a \big) \,
d E = 0 \\
            & \lim_{\lambda,\eta\to 0 }\,\int_I \delta\Big( \Big| \cos\big(\sqrt{E+i\eta} \,
L_e(\lambda,\cdot)\big) \notag \\ 
& \mkern70mu  + \frac{\sin\big(\sqrt{E+i\eta}\,
L_e(\lambda,\cdot)\big)}{\sqrt{E+i\eta}} R^+_e(0;E+i\eta,\lambda,\cdot)
            \Big|^2 ,a \Big)\;
            dE  = 0
\end{align}
for all $ a \in (0,1/2] $. By \cite[Lemma~D.4]{ASW05}
this implies that both random variables in \eqref{eq:ImMod} are almost
surely constant for Lebesgue-almost all $ E \in I $. Since
they are identically distributed for all $ e \in \mathcal{E} $,
the constants $ \mathcal{I} $ and $ \mathcal{M} $ are independent of $ e $ .
\qed
\end{nummer}
\end{proof}
The explicit expression \eqref{eq:specSol} shows that $ \sin\big(\sqrt{E} L\big)/
\sqrt{E} \neq 0 $ for all $ E \in \sigma_{\rm ac}(-\Delta_{\T{}}) $.
Therefore, \eqref{eq:ImMod} asserts that the $ R_e $-marginals of $
\nu^E $ are supported on
the intersection of a line with a circle, that is, on
at most two points.
Next, we show that this support contains only one point which
coincides with $ R_0(0,E+i0,0) $.
\begin{lemma}\label{lemma:onep}
Assume the situation of Lemma~\ref{lem:twop}. Then for
Lebesgue-almost all $ E \in I $:
\begin{indentnummer}
\item there exists $ \Phi \in \C $ with $ \Im \Phi \geq 0
$ such that for all $ e\in \mathcal{E} $
\begin{equation}
R_e =  \Phi \qquad \mbox{$ \nu^E $-almost surely.}
\end{equation}
\item $ \Phi = R_0(0,E+i0,0) $.
\end{indentnummer}
\end{lemma}
\begin{proof}
\begin{nummer}
\item By Lemma~\ref{lem:twop} there exists $ \Phi^\pm \in
\C $ with $ \Im \Phi^\pm \geq 0 $
such that the $ R_e $-marginal of $ \nu^E $ is supported on
$ \{ \Phi^+, \Phi^- \} $ for all $ e \in \mathcal{E} $.
Suppose  that $ \Phi^+\neq \Phi^-$.
Then the distribution of $ \sum_{f\in \mathcal{N}^+_e} R_{f}
$ is supported on at least three points. This follows by explicitly
identifying three points $ K \Phi^\pm $ and $ (K-1)
\,\Phi^+ +  \Phi^- $ in the support. But this contradicts the
limiting
recursion relation~\eqref{eq:reclim} since then the
distribution of the left side is supported contains at least three points
but the
distribution of the right side on at most two points in its support.
\item
Equation~\eqref{eq:reclim} with   $ \Phi $  substituted for all
$ R_e $, and $ L_e = L $ for all edges $e$,  is quadratic
in $ \Phi $.    For Lebesgue almost all $ E \in
\sigma_{\rm ac}(-\Delta_{\T{}}) $ this equation has a complex
non-real solution, and in this case $ R^+_0(0,E+i0,0) $ is
its only solution in the upper half plane.
\qed
\end{nummer}
\end{proof}


\subsection{Section summary}

Let us now note that the above lemmas imply the two theorems
stated in this section: \\

\noindent {\it Proof of Theorem~\ref{thm:meas}.\/}     ~
Lemmas~\ref{lem:twop}
and \ref{lemma:onep}  jointly imply  Theorem~\ref{thm:meas}.  \qed \\

\noindent   {\it Proof of  Theorem~\ref{thm:stoch}.\/}  ~
As is discussed in \cite{ASW05}, an application of Fatou's lemma yields \eqref{eq:stoch} from \eqref{eq:disconv}.
  \qed \\

\section{Extensions} \label{sec:extensions}

\subsection{More general vertex conditions}  
A variety of boundary conditions other  than \eqref{eq:Kirch} lead to self-adjoint
Laplacians on metric graphs \cite{KosSch99,Kuch04}.  Of those, the 
argument presented here can be readily extended to the class of  
symmetric BC.
These  require at each vertex:
\begin{enumerate}
\item for some fixed $ \beta \in [0,\pi] $ the following  is common 
to all the edges $ e $ adjacent to the vertex    
\begin{equation} 
 \cos(\beta) \, \psi + \sin(\beta) \; n_{e} \cdot \nabla \, \psi 
\end{equation}
with  $  n_{e} \cdot \nabla $ the \emph{outward} derivative, 
\item for some  fixed $ \alpha \in [0,\pi] $ the sums over all edges
	adjacent to the vertex satisfies
\begin{equation} 
 \cos(\alpha) \  \sum_{e} \psi_e  -  \sin(\alpha) \  \sum_{e} n_{e} \cdot \nabla \, \psi  \ = \ 0 \, ,
\end{equation} 
	\end{enumerate}
The symmetric class includes the Kirchhoff BC \eqref{eq:Kirch}, for which $ \beta = 0 $ and 
$\alpha = \pi/2 $.     

Our analysis extends to  the general symmetric BC through a rotation which mixes 
the function $\psi^+$ and its derivative, where $\psi^+$ is defined as below  
 Theorem~\ref{prop:unique} with the present boundary conditions.   We  denote: 
\begin{equation}\label{eq:defpsit}
	 {\widetilde\psi^+}(x;z|0) \  
	:=\   \cot(\beta) \, \psi^+(x;z|0) + \frac{\partial}{\partial x}\psi^+(x;z|0)  
	\, = - \frac{\psi^+(x;z|0)}{{\widetilde R}^+(x;z)}  \,  ,  
\end{equation}
and correspondingly
\begin{equation}
	{\widetilde R}^+(x;z) :=  - \left[ \cot(\beta) +  R^+(x;z) \right]^{-1} \, .  
\end{equation}  
Under the above boundary conditions it is  the function  ${\widetilde\psi^+}(x;z|0) $ 
which takes a common value among the forward edges of any vertex. 
The current can be expressed in terms of the 'rotated' quantities as 
\begin{equation} 
 J^+(x;z|0) \equiv  \big|{ \psi^+}(x;z|0)  \big|^2  \Im {  R}^+(x;z)  
\ = \ \big|{\widetilde\psi^+}(x;z|0)  \big|^2  \Im {\widetilde R}^+(x;z)  \, . 
\end{equation}     

The argument, as it is outlined in Section~\ref{subsec:outline}, applies verbatim with   
$\psi^+(x;z|0) $  and $ R^+(x;z)$ replaced by 
${\widetilde\psi^+}(x;z|0) $  and $ {\widetilde R}^+(x;z)$.   In this context the relevant  Lyapunov exponent is 
\begin{equation}
		{ \widetilde \gamma_\lambda}(z) \ := \  - \mathbb{E}\left[ \log \left( \sqrt{ K} \,
			\frac{\big| {\widetilde \psi_f^+}(0;z, \lambda, \cdot |0) \big|}{\big| {\widetilde \psi_0^+}(0;z, \lambda, \cdot |0) \big|}\right) \right] \, , 
	\end{equation}   
where $f$ is an arbitrary edge forward to the edge emanating from the root. 
It follows from  \eqref{eq:defpsit} that the above expression 
with ${\widetilde\psi^+}  $ yields the same value as with ${\psi^+}  $, i.e., 
${ \widetilde \gamma_\lambda}(z) = {  \gamma_\lambda}(z)$, the latter being defined by \eqref{eq:defgamma}.  

\subsection { Tree graphs with decorations } 

 By gluing a copy of a finite metric graph $ \G $ to every vertex of the tree $ \T $  
 one obtains a metric graph $ \T \lhd \G $ 
which is refered to as a \emph{decorated}  tree.  
The Laplacian $ -\Delta_{\T \lhd \G} $ is rendered self-adjoint by imposing, 
for example, Kichhoff BC.   Such decorations provide a mechanism for the 
creation of gaps in the spectrum \cite{ScAi00,Kuch05}.    

The strategy presented here allows to establish the stability of the 
AC spectrum under random deformations of a (uniformly) decorated tree 
even if $ \G $ has loops.  In deriving the fluctuation bounds in this case,  
in the sum on the left side of \eqref{eq:recur} one  may omit the terms 
$ \Im R^+_f $ which correspond to directions $ f $ into the decorating parts.  
These terms vanish  
for real $ z $ since the finite graph $ \G $ does not conduct current to infinity.  
 

\appendix

\section{Appendix:~More on the Weyl-Titchmarsh function on tree
graphs}\label{app:more}
%
This appendix is devoted to the WT functions~$ R^\pm $
on general metric tree graphs $ \T $, presented in
Definition~\ref{def:WT}.  We start by proving
Theorem~\ref{prop:unique} on which the definition relies.
Basic properties of $ R^\pm $ are the topic of the second subsection.
The third subsection deals with the Green function on $ \T $ and its
relation to the WT functions.
\subsection{ Uniqueness of square-integrable solutions on graphs with
a dangling end} We will now give a proof of Theorem~\ref{prop:unique}.
\begin{proof} \begin{nummer}
\item That
there is at least one non identically vanishing function
in the kernel of the operator $- \Delta_{\G,u}^* - z $ can be seen by
elongating the dangling edge beyond $ u $ thereby creating a backward extension $ \G' \supset \G $ .
We set
\begin{equation}\label{eq:sol}
           \psi(x;z) = \left( -\Delta_{\G'} -z \right)^{-1} \varphi(x)
\end{equation}
where $ \varphi $ is some non identically vanishing function compactly supported on the elongation of the edge containing $ u $
and $  -\Delta_{\G'} $ is a
\emph{self-adjoint}
Laplacian on $ \G' $.
This function~(\ref{eq:sol}) does
not vanish identically on $ \G $, since otherwise (by (\ref{eq:explsol}) below)
it would be identically zero on the whole edge containing $ u $ and
the support of $ \varphi $.\\
Suppose now there is another solution
which is linearly independent of $ \psi(\cdot ; z) $.
Since the solution space on the edge adjacent to $ u $ is two
dimensional, one can linearly combine them to satisfy
a self-adjoint BC (cf.\ \eqref{eq:root}) at $ u $.
Thereby one produces an eigenfunction of a self-adjoint Laplacian $ -\Delta_\G $ with eigenvalue $ z \in \C^+ $.
This contradicts the self-adjointness.
 \item In fact, more generally for all $ x \in \G $ which
disconnect the graph, we have that $ \cos(\alpha) \, \psi(x;z)
- \sin(\alpha) \,  \psi'(x;z) \neq 0 $
for all $ \alpha \in [0, \pi) $. Otherwise one would have
found a square-integrable, non-trivial eigenfunction with eigenvalue
$ z \in \C^+ $ of a restriction of $ -{\Delta}_{\G,u}^* $
to functions on that
disconnected piece, which does not
contain $ u $. Since this Laplacian is rendered self-adjoint by
imposing $ \alpha $-BC at $ x $,
this is a contradiction.
\item This is an immediate consequence of \eqref{eq:sol}.
\qed
\end{nummer}
\end{proof}

\subsection{Basic properties of the Weyl-Titchmarsh functions}
Following are some properties of $ R^+(x;z ) $ which are of relevance in the main part of the paper.
Similar statements apply to $ R^- $, with proofs differing only in
the notation.
\begin{theorem} \label{thm:profR}	The WT function $ R^+(x;z ) $ has the following properties:
\begin{indentnummer}
	\item  $ R^+(x;\cdot ):\C^+ \to  \C^+ $ is analytic for fixed $ x $.
	\item For each $ e \in \mathcal{E} $ and all $ z \in \mathbb{C}^+$:
	\begin{equation}\label{eq:boundR}
	 \big| R^+_e(0;z) \big| \leq  \frac{2 \sqrt{|z|}}{1-\exp\left(-2 L_{e} \Im \sqrt{z}\right)},
	\end{equation}
	and
	$ |  R^+_e(L_e;z) | \leq 2 K  \sqrt{|z|} \left[1-\exp\left(-2 L_{e} \Im \sqrt{z}\right)\right]^{-1} $
	due to \eqref{eq:add}.
	\item Along any edge $ e $ the function obeys the
	Riccati equation \eqref{eq:Riccati}. In particular its values
	are related by M\"obius transformations:
	\begin{equation}\label{eq:Moebius}
                  R^+_e\big(l;z\big) =
	\frac{R^+_e\big(0;z\big)\cos\big(\sqrt{z} l \big) - \sqrt{z}
			\sin\big(\sqrt{z} l \big)
                   }{\cos\big(\sqrt{z} l \big) +  R^+_e\big(0;z\big)
	\sin\big(\sqrt{z} l\big)/\sqrt{z} }
	\end{equation}
	for all $ l \in [0, L_e] $ and $ z \in \C^+ $.
	\item Equipping the space $ [L_{\min}, \infty)^\mathcal{E} $ with the uniform topology, $ R_0^+(0;z) $ is a continuous function
	of $ \{L_e\}_{e \in \mathcal{E} } \in [L_{\min}, \infty)^\mathcal{E} $ for all $ z \in \C^+ $.
	\end{indentnummer}
\end{theorem}
\begin{proof} \begin{nummer}
	\item
	The first assertion follows from the analyticity of $ \psi^+(x;z | 0 ) $ and of its deriative (cf.\ Theorem~\ref{prop:unique}).
	The Herglotz-Nevanlinna property is a consequence of~\eqref{eq:RGreen}.
	\item This is an immediate consequence of \eqref{def:w} and Lemma~\ref{lemma:w}(i) below.
	\item This assertion follows from the fact that $ \psi_e^+ (\cdot;z |0) \in {\rm H}^2[0,L_e] $ is a solution
         of the free Schr\"odinger equation $ -\psi'' = z  \psi $
  	on the interval $ [0,L_e] $ which, using the boundary conditions at $ l = 0 $, may be written as
      \begin{equation}\label{eq:explsol}
	 \frac{\psi_e^+(l;z | 0 )}{\psi_e^+(0;z |0)} =  \cos(\sqrt{z} l)  + R^+_e(0;z)\,
	\frac{\sin(\sqrt{z} l)}{\sqrt{z}}
\end{equation}
for all $ l \in [0,L_e] $.
	\item Suppose the metric tree $ \T $ is finite and has only $ N $ generations, i.e., 
the number of edges connecting any edge to the root is at most $ N $. In this case, the continuity of $ R_0(0;z) $ follows from the explicit
	evolution equations \eqref{eq:Moebius} and \eqref{eq:add}. Lemma~\ref{lemma:w}(iii) below shows that
	$ R_0(0;z) $ may be uniformly
	approximated by its values on a finite tree provided $ \Im \sqrt{z} $ is large enough. Hence $ R_0(0;z) $
	is continuous for those $ z \in \C^+ $.
	Since $ R_0(0;z) $ is analytic in $ z \in \C^+ $, this implies continuity for all $ z \in \C^+ $.  \qed
\end{nummer}
\end{proof}

\begin{remark}\label{rem:bound}
Another immediate consequence of \eqref{eq:explsol} and its analog
with $ 0 $ and $ L_e $ interchanged, is the bound
\begin{equation}\label{eq:boundsol}
e^{-\sqrt{|z|} L_e}\left( 1 + \frac{ |R^+_e(L_e;z)|}{\sqrt{|z|}}
\right)^{-1} \leq
	\left|\frac{\psi_e^+(L_e;z |0)}{\psi_e^+(0;z |0)}\right| \leq
e^{\sqrt{|z|} L_e} \left( 1 + \frac{ |R^+_e(0;z)|}{\sqrt{|z|}} \right)
\end{equation}
which shows that $ \psi^+(\cdot;z | 0) $ de- or increases at most
exponentially on any edge $ e $.
\end{remark}
Instead of the WT function $ R^+ $, it is sometimes more convenient to consider its transform
\begin{equation}\label{def:w}
	m(x;z) := \frac{R^+(x;z) - i \sqrt{z}}{R^+(x;z) + i \sqrt{z}}
\end{equation}
which takes values in the complex unit disk. The evolution on the edges takes a particularly simple form for $ m $.
In fact, from \eqref{eq:Moebius} and \eqref{eq:add} one obtains
\begin{equation}\label{eq:w1}
 m_e(0;z) = e^{2i \sqrt{z} L_e}  m_e(L_e;z), \quad \mbox{and}\quad
	m_e(L_e;z) = g\Bigg( \sum_{f \in \mathcal{N}^+_e} \frac{1+m_f(0;z)}{1-m_f(0;z)} \Bigg),
\end{equation}
where $ g(\zeta) := \frac{\zeta -1}{\zeta + 1} $. The next lemma collects some facts which are used in the proof of
Theorem~\ref{thm:profR}.
\begin{lemma}\label{lemma:w}
Let $ z \in \C^+ $ and assume $ L_e \geq L_{\min} > 0 $ for all $ e \in \mathcal{E} $. Then $ m(x;z)  $
has the following properties:
\begin{indentnummer}
\item It satisfies: $  | m_e(0;z) | \leq \exp\left( - 2 L_{e} \Im \sqrt{z}\right) $.
\item At the root the dependence on a particular value $ m_e(L_e;z) $ is uniformly exponential in the sense that there exists a constant
	$ c < \infty $ such that for all $ \Im \sqrt{z} $ sufficiently large:
\begin{equation}
	 \left| \frac{\partial  m_0(0;z)}{\partial m_e(L_e;z)}\right| \leq c^N \exp\left(-2 N L_{\min} \Im \sqrt{z}\right)
\end{equation}
where $ N $ is the number of vertices between the edge $ e $ to the root.
\item Let $ m_0(0;z) $ and $ {\widetilde m_0}(0;z) $ correspond to metric tree graphs $ \T $ and $ \widetilde \T $ which coincide up to the
$N$th generation.
Then for all $ \Im \sqrt{z}  $ sufficiently large:
\begin{equation}
	\left| m_0(0;z) - {\widetilde m_0}(0;z) \right| \leq 2 K^{N+1}  c^N \exp\left(-2 N L_{\min} \Im \sqrt{z} \right).
\end{equation}
\end{indentnummer}
\end{lemma}
\begin{proof}
	\begin{nummer}
	\item This is an immediate consequence of the first evolution equation \eqref{eq:w1}.
	\item Using the chain rule this can be traced back to a straightforward differentiation of the equations \eqref{eq:w1}.
	The edge and vertex terms are subsequently bounded with the help of~(i).
	\item We expand the difference into a telescopic sum of $ K^{N+1} $ differences and use both~(ii) and the fact that
	the values of $ m(\cdot ; z )$ and $ {\widetilde m}(\cdot ; z ) $ on the $ K^{N+1} $ leaves in the $ N $th generation
	differ at most by a complex number of modulus $ 2 $. \qed
	\end{nummer}
\end{proof}
\subsection{The Green function on a tree graph}
Analogously to one dimension \cite{CoLe55,CaLa90}, the Green function
of the Laplacian $ -\Delta_\T $ on a metric tree graph $ \T $
can be constructed using two non-vanishing square-integrable
functions. In fact, the following lemma is straightforward.
\begin{lemma}
The Green function $ G_\T(u,x;z) $ of the Laplacian $ -\Delta_\T $ can be expressed as
\begin{equation}\label{eq:Green1}
           G_\T(u,x;z) = \frac{\big(\psi^+ \wedge
\psi^-\big)(u,x;z | v)}{W(\psi^+,\psi^-)(u; z |v )},
\end{equation}
independently of $ v $, as long as $ v \in \T^+_u \cap \T^+_x $. Here
\begin{equation}
	\big(\psi^+ \wedge\psi^-\big)(u,x;z |v ) := \left\{
\begin{array}{l@{$\quad\mbox{for}\quad$}l}
                   \psi^+(u;z | 0)\, \psi^-(x;z |v ) & x \in \T^-_u \\
                   \psi^-(u;z |v )\, \psi^+(x;z| 0) & x \in \T^+_u \, .
                           \end{array} \right.
\end{equation}
and $ W(\psi^+,\psi^-) := \psi^+
\big({\partial \psi^-/\partial x}\big)  -  \big({\partial \psi^+/\partial x} \big) \psi^- $ is
the Wronskian.
\end{lemma}
\begin{remarks}
\begin{nummer}
\item
The Wronskian is constant along any edge in $ \T_v^- $. In
particular, this implies
that $ W(\psi^+,\psi^-) \neq 0 $, since otherwise one could linarly combine $ \psi^\pm $ to a
square-integrable solution of $(-\Delta_\T - z) \, \psi = 0 $ on the
whole tree.
\item
The right side of~\eqref{eq:Green1} defines an integral
         kernel of the resolvent $ (-\Delta_\T - z)^{-1} $ which is
	jointly continuous in $ (u,x) $.
	\item Setting $ f_v^\pm(\cdot;z|\cdot ) :=
\psi^\pm(\cdot;z|\cdot)/\psi^\pm(v;z |\cdot) $, where $ v $ is any point
	on the same edge as $ u $, the Green function~\eqref{eq:Green1}
	can be rewritten in terms of WT functions:
	\begin{equation}\label{eq:ox}
	 	G_\T(u,x;z) = - \frac{\big(f_v^+\wedge
f_v^-\big)(u,x;z)}{R^+(v;z) + R^-(v;z)}.\qquad\qquad\;
	\end{equation}
	In particular, for $ u = x = v $, we obtain~\eqref{eq:Green}.
Moreover, at the root,
	we obtain
	\begin{equation}
		G_\T(0,0;z) = \big( \cot \alpha - R^+(0;z)
\big)^{-1}, \qquad \alpha \neq 0,
	\end{equation}
	because $ R^-(0;z) = - \cot \alpha $ due to the BC \eqref{eq:root}.
\end{nummer}
\end{remarks}

For a self-adjoint Sturm-Liouville, or more specifically,
Schr\"odinger  operator on the half-line, the WT function at the
origin
allows one to reconstruct the spectral measure and therefore contains
all spectral information \cite{CoLe55,CaLa90}. Generally, this fails
to hold for operators on tree graphs.
However, the AC spectrum of $ - \Delta_\T $ can still be detected by
the boundary value of $ R^+(0;z) $:
\begin{theorem}\label{thm:spec}
The AC spectrum $ \sigma_{\rm ac}(-\Delta_\T) $ of the Laplacian
on a rooted
metric tree graph $ \T $ is concentrated on the set
\begin{equation}\label{eq:spec}
  \left\{ E \in \R \, : \, 0 < \Im R^+_0(0;E+i0) < \infty \right\}.
\end{equation}
\end{theorem}
\begin{proof}
	Pick any edge $ e \in \mathcal{E} $ and let $ \phi $ be a
compactly supported function on  $e$.
	A straightforward but tedious computation using \eqref{eq:ox} shows that
for Lebesgue-almost all $ E \in \R $
	the AC density of the spectral measure associted with $ \phi $
	is given by
	\begin{multline}\label{eq:specm}
	\lim_{\eta \downarrow 0} \;\, \Im \left\langle \phi , \,
	\left( - \Delta_\T - E - i \eta \right)^{-1} \phi \right\rangle	 \\
	=  \frac{\Im R^+_e(0;E+i0)\; g_{\phi}^-(E) + \Im
R^-_e(0;E+i0)\; g_{\phi}^+(E) }{|R^+_e(0;E+i0) + R^-_e(0;E+i0)|^2},
	\end{multline}
	for Lebesgue-almost every $ E \in \R $, where
	\begin{equation}
	 g_{\phi}^\pm(E) := \big| \big\langle \phi, \Re
f^\pm\big(\cdot ;E\big) \big\rangle\big|^2
	+ \big| \big\langle \phi, \Im f^\pm\big(\cdot ;E\big)
\big\rangle\big|^2
	\end{equation}
	and $ f^\pm\big(\cdot ;E\big) $
	is the solution of the Schr\"odinger equation $ \left(-
\Delta_\T - E \right) f = 0 $,
	which satisfies $ \big(d f_{f}^\pm/dx\big)\big(0 ;E \big) =
\pm R^\pm_{f}(0;E+i0) $ at every
	edge $ f $ and is normalized to
	$ f^\pm_e\big(0 ;E\big) = 1 $.
	By the current conservation \eqref{eq:defcurrent} along each
edge and the positivity and additivity of the current at
	each vertex, we have
	\begin{equation}
		\Im R^+_{e}(0;E+i0) \leq \left| f^+_0\big(0 ;E\big)
\right|^2 \, \Im   R^+_0(0;E+i0)
	\end{equation}
	for Lebesgue-almost all $ E \in \R $. Similarly,
	by tracing the current flow on the backward tree emanating
from $ e $ and using $ \Im   R^-_0(0;E+i0) = 0 $,
  	we obtain for Lebesgue-almost all $ E \in \R $
	\begin{equation}
		\Im R^-_e(0;E+i0) = \sum_{f} \left| f^-_{f}\big(0
;E\big) \right|^2 \Im R^+_{f}(0;E+i0)
	\end{equation}
	where the sum extends over all edges $ f\neq e $, which have
the same distance to the root as~$ e $.\\
	From the above considerations we conclude that for any $ e $
and Lebesgue-almost all $ E \in \R $ if $ \Im   R^+_0(0;E+i0) = 0 $ then
	{\it 1.}~ $ \Im R^+_{e}(0;E+i0) = 0 $, and {\it 2.}~ $ \Im  R^-_e(0;E+i0) = 0 $.
	But this shows that $ \sigma_{\rm ac}(-\Delta_\T) $ is indeed
concentrated on the set in \eqref{eq:spec}.
	 \qed
\end{proof}

\begin{acknowledgement}

We are grateful to Uzy Smilansky for stimulating discussions
concerning quantum graphs.  We would also like to thank for
the gracious hospitality
enjoyed at the Weizmann Institute (MA) and the Department of
Mathematics at UC Davis (SW).   This work was supported in parts by
the Einstein Center for Theoretical Physics and the Minerva Center
for Nonlinear Physics at the Weizmann Institute, by the US National
Science Foundation, and by the Deutsche Forschungsgemeinschaft.

\end{acknowledgement} %

\end{document}